\documentclass[a4paper,12pt]{article}
\usepackage{jheppub} 
\usepackage{amsfonts}
\usepackage{amssymb}
\usepackage{multicol}
\usepackage{graphicx}
\usepackage{float}
\usepackage{caption}
\usepackage{xcolor}
\usepackage[utf8]{inputenc}
\usepackage{amsmath}
\usepackage{bm}
\usepackage{orcidlink}

\newcommand{\Tr}{\mathrm{Tr}}

\newcommand{\Rsun}{R_{\odot}}
\newcommand{\QFI}{\mathcal{H}}
\newcommand{\FI}{\mathcal{F}}
\newcommand{\Diss}{\mathcal{D}}
\newcommand{\ident}{\mathcal{I}}

\definecolor{blue}{RGB}{0,65,180}
\newif\ifrevmark
\ifdefined\cleanbuild\revmarkfalse\else\revmarktrue\fi
\newcommand{\revision}[1]{\ifrevmark\textcolor{blue}{#1}\else#1\fi}

\title{Conditional multiparameter thermometry with neutrino oscillations: MSW energy diversity and a plasma-noise benchmark}
\author[1]{Jilali Loulijat\orcidlink{0009-0005-2517-7878}}
\author[2,3]{Abdallah Slaoui\orcidlink{0000-0002-5284-3240}}\emailAdd{abdallah.slaoui@um5s.net.ma}
\author[2,3,4]{El Hassan Saidi\orcidlink{0000-0003-4039-6700}}
\author[1]{Mohamed Gouighri\orcidlink{0000-0002-9551-0251}}
\author[5,6]{Berihu Teklu\orcidlink{0000-0001-9280-533X}}
\emailAdd{berihu.gebrehiwot@ku.ac.ae}
\affiliation[1]{LPMS, Faculty of Sciences, Ibn Tofail University, Kenitra, Morocco.}
\affiliation[2]{LPHE--Modeling and Simulation, Faculty of Sciences, Mohammed V University in Rabat, Rabat, Morocco.}
\affiliation[3]{CPM, Faculty of Sciences, Mohammed V University in Rabat, Rabat, Morocco.}
\affiliation[4]{College of Physical and Chemical Sciences, Hassan II Academy of Sciences and Technology, Rabat, Morocco.}
\affiliation[5]{College of Computing and Mathematical Sciences, Department of Applied Mathematics and Sciences, Khalifa University of Science and Technology, 127788 Abu Dhabi, United Arab Emirates.}
\affiliation[6]{KU Research Center for Intelligent Computing, Networks and Sensing (ICONS), Khalifa University of Science and Technology, 127788, Abu Dhabi, United Arab Emirates.}

\abstract{We address temperature-parameter inference in a restricted two-flavor neutrino dephasing model with fixed mean electron densities. When two zone temperatures enter the propagated state via a single accumulated dephasing exponent, the single-energy quantum Fisher information matrix has rank at most 1. Independent, energy-resolved samples restore regular local identifiability when their thermal-sensitivity vectors are noncollinear. Matter-dependent mixing provides distinct weights in an illustrative two-zone model. A separate two-probe benchmark shows how complementary Bell sectors can distinguish temperature from spatial noise correlations. An electron-only Debye-cell estimate gives a conditional longitudinal quantum signal-to-noise ratio of approximately $3.1\times10^{-13}$ at 10 MeV for a uniform central-density medium over one solar radius. This estimate is not a microscopic upper bound or a detector sensitivity forecast. Complete phase averaging removes temperature information carried solely by longitudinal coherence damping; to retain solar sensitivity, a separate calculation of population, refractive, or production effects is required. The results distinguish conditional state identifiability from operational temperature inference.}


\begin{document}
\maketitle
\flushbottom
\section{Introduction}
Quantum thermometry asks how accurately a temperature parameter can be inferred from the state of a microscopic probe. Because temperature is not represented by a unique Hermitian observable of the probe, the problem is naturally formulated within quantum estimation theory. The quantum Fisher information (QFI) determines the ultimate single-parameter Cram\'er--Rao precision, while the QFI matrix (QFIM) exposes correlations and identifiability in multiparameter problems \cite{Paris2009,Correa2015}. In open-system thermometry, the probe need not fully thermalize with the sample. In particular, a qubit may acquire temperature information through dephasing, so that coherence loss becomes a sensing resource rather than merely a nuisance \cite{Razavian2019,Sehdaran2019}.

Recent work has broadened this perspective in several directions. 
Two-qubit probes distinguish local from common reservoirs and show that entanglement, coherence, and bath correlations contribute differently in different interrogation-time regimes \cite{Gebbia2020}. Coherence generation can broaden or enhance thermometric sensitivity \cite{Ullah2023}. More recently, two-qubit thermometry under local and global pure-dephasing environments has shown that bath-induced correlations and reservoir spectral properties can substantially reshape the attainable temperature sensitivity, with common-bath configurations being advantageous in particular interrogation-time regimes \cite{Hminat2026Hyperon}. Non-Markovian thermometry has also been extended to relativistic settings, where an ancillary Unruh--DeWitt detector can mediate the transfer of temperature-dependent information to the probe coherence and modify the achievable quantum Fisher information \cite{Hminat2026AdS}. A complementary spectral viewpoint emphasizes that a probe's energy gaps determine the temperature window to which it is sensitive~\cite{Correa2015}. Multiparameter thermometry has also been considered explicitly in protocols designed to infer two temperatures \cite{VermaCosta2024}.

Earlier work by Tamascelli \emph{et al.} explicitly examined thermometric probing beyond pure dephasing \cite{Tamascelli2020}. Multiparameter compatibility and the distinction between state information and feasible readout have also been studied in driven-dissipative Kerr resonators \cite{AsjadTekluParis2023}. These results motivate checking both the physical coupling and attainable measurement, without transferring conclusions between different bath models.

Neutrino oscillations provide an unusual setting in which several of these ingredients occur naturally. Flavor states are coherent superpositions of mass states, and in the two-flavor approximation, the relevant Hilbert space is two-dimensional. Matter modifies the Hamiltonian through the Wolfenstein potential and the MSW mechanism \cite{Wolfenstein1978,MikheyevSmirnov1985}. Stochastic-matter studies established long ago that density fluctuations can alter flavor conversion \cite{LoretiBalantekin1994,Nunokawa1996,BurgessMichaud1997}. More recently, quantum-estimation methods have been applied directly to neutrino oscillation parameters, clarifying both the role of off-diagonal QFIM elements and the gap between the information encoded in the state and that accessible to flavor measurements \cite{Frugiuele2026,Ignoti2025CP,Huang2026}.\par

Recently, we studied how dephasing changes entanglement of formation, quantum discord, and local quantum uncertainty in an effective two-mode description of two-flavor neutrino oscillations \cite{LoulijatQuantumness2026}. That analysis concerns the survival of quantum correlations. Related investigations have also examined the dynamics of quantum-information resources in two-flavor neutrino oscillations under correlated dephasing, including quantum coherence, entanglement, local quantum Fisher information, and Bell nonlocality in both Markovian and non-Markovian regimes \cite{ElBouzaidi2025}. These studies primarily characterize the persistence, degradation, and possible revival of quantum correlations in the presence of environmental noise.\par

The present work instead asks whether a specified temperature-dependent channel encodes identifiable thermal parameters. Correlation measures alone do not establish attainable precision in temperature. Two-flavor quantum estimation and flavor-readout limitations were already investigated by Nogueira \emph{et al.} \cite{Nogueira2016}. Recent operational work distinguishes spectral rank restoration from detector information \cite{Lu2026}. Our related preprint addresses neutrino decoherence, quantum resources, and multiparameter incompatibility \cite{LoulijatTeklu2026}. The present study instead concerns conditional temperature-profile encoding. Thus, the general principles of energy-resolved rank restoration and state-versus-readout information are prior background, rather than new claims here. These developments motivate neutrino thermometry, but they also impose a crucial physical consistency condition. The ordinary MSW potential
\begin{equation}
V(r)=\sqrt{2}G_F n_e(r)
\end{equation}
is primarily a probe of the mean electron density. Temperature can be inferred only if a separate, microscopically controlled temperature dependence enters the neutrino state. In the present framework, this information is carried by the two-point spectrum of electron-density fluctuations, $S_{nn}(\bm q,\omega; T)$. A freely adjustable phenomenological decoherence rate is therefore insufficient: the mean potential, fluctuation amplitude, correlation time, and finite-temperature response must be consistent with the same medium.

The Sun is a useful benchmark because its density and temperature profiles are constrained by standard solar models \cite{Vinyoles2017}, its $^8$B neutrino interaction rate has been measured by Borexino \cite{Borexino2020}, and inverse analyses have already shown that flavor data can constrain the solar electron-density profile \cite{LaberSmith2024}. Current global oscillation parameters are also known with sufficient accuracy to make the two-flavor solar sector a controlled starting point \cite{NuFIT2024}. At the same time, the heuristic electron-only Debye-cell construction used below gives an extremely small longitudinal neutrino decoherence scale. This creates an instructive tension: the information geometry may be nontrivial even when the absolute thermal imprint is experimentally negligible.

The central question of this work is therefore not simply whether a QFI for temperature can be written. We ask instead: \emph{what thermal combinations are identifiable from neutrino propagation, how can MSW energy dependence lift degeneracies, what additional information is supplied by two probes in local or correlated baths, and do these information-theoretic advantages survive the microscopic solar scale?}

Our main results are fourfold. First, the MSW transformation rotates the density-noise coupling between longitudinal and transverse directions, producing an energy-dependent sensing geometry. Second, for a two-zone temperature model, a single energy in the longitudinal dephasing limit yields a rank-one QFIM for $(T_1, T_2)$; separate temperature inference is therefore structurally impossible within that channel. Third, different energies generate different sensitivity directions and can make the combined QFIM full rank. Fourth, common and differential bath correlations can be separated with two-probe Bell sectors, but the resulting metrological structure does not alter the microscopic weak-interaction scale. The conclusion is thus deliberately conservative: energy diversity can solve an \emph{identifiability} problem, but it does not solve the Sun's \emph{detectability} problem. This detectability statement is restricted to the explicitly analyzed longitudinal dephasing benchmark. It is not a general solar bound: complete unobserved phase averaging removes the temperature imprint carried solely by longitudinal coherence damping, while transverse transitions, temperature-dependent refraction, and production effects require separate calculations. For the high-energy-physics reader, the practical message is twofold: (i) any proposal to extract medium temperature from neutrino propagation must first show that the thermal parameters enter through more than one independent sensitivity direction, and (ii) the absolute size of the thermal imprint must be fixed by the same microscopic weak-interaction physics that sets the MSW potential, not by a free decoherence parameter. This formulation belongs to the broader framework of quantum metrology and sensing \cite{Giovannetti2011,Degen2017}. Reviews of quantum thermometry distinguish equilibrated probes from dynamical protocols \cite{Mehboudi2019}. Local quantum thermal susceptibility characterizes temperature information in reduced equilibrium states \cite{DePasquale2016}; the propagating neutrino considered here is instead a nonequilibrium probe, so equilibrium response bounds cannot be transferred without checking their assumptions.

\section{Quantum-estimation framework}
The statistical-distance formulation provides the geometric foundation of quantum Fisher information \cite{BraunsteinCaves1994}. Here, the QFIM is a local state-level benchmark; its operational use also requires specifying accessible measurements and counting independent probes and interrogation time.

Let $\rho_{\bm\lambda}$ be the neutrino state after propagation through a medium characterized by parameters $\bm\lambda=(\lambda_1,\lambda_2,\ldots)$.  The symmetric logarithmic derivatives $L_\mu$ satisfy
\begin{equation}
\partial_{\mu}\rho=\frac{1}{2}\left(L_\mu\rho+\rho L_\mu\right),
\end{equation}
and the QFIM is
\begin{equation}
\QFI_{\mu\nu}=\frac{1}{2}\Tr\left[\rho\left(L_\mu L_\nu+L_\nu L_\mu\right)\right].
\end{equation}
For $M$ statistically independent realizations, the covariance matrix of any locally unbiased estimator obeys
\begin{equation}
\mathrm{Cov}(\hat{\bm\lambda})\geq \frac{1}{M}\QFI^{-1},
\label{eq:qcrb}
\end{equation}
under the usual regularity and local-unbiasedness assumptions on an identifiable parameter subspace. The SLD matrix inequality remains a lower bound even if equality cannot be attained simultaneously. Attainability is a separate issue \cite{Ragy2016,Liu2020}. Nonsingularity characterizes regular first-order local identifiability in the stated coordinates.

For one temperature $T$, we use the dimensionless quantum signal-to-noise ratio (QSNR)
\begin{equation}
Q_T=T^2\QFI_{TT},\qquad
\frac{\delta T}{T}\geq \frac{1}{\sqrt{M Q_T}}.
\label{eq:qsnr}
\end{equation}
Here, all other unknown parameters are fixed. With nuisance parameters, the corresponding effective quantity is $Q_T^{\rm eff}=T^2/[(\QFI^{-1})_{TT}]$ when the full inverse exists. Singular models require restriction to identifiable combinations.
For an actual measurement with probabilities $p_k(\bm\lambda)$, the classical Fisher matrix is
\begin{equation}
\FI_{\mu\nu}=\sum_k\frac{\partial_\mu p_k\,\partial_\nu p_k}{p_k}.
\label{eq:cfi}
\end{equation}
The distinction between QFI and flavor-accessible FI is especially important for neutrinos because arbitrary optimal POVMs are not currently experimentally available \cite{Frugiuele2026,Huang2026}.

\section{MSW neutrino as an open-system thermometer}

\subsection{Mean Hamiltonian}

We use $\hbar=c=k_B=1$ in the formal Hamiltonian and spectral equations. Temperatures quoted in kelvin correspond to energies $k_BT$; for numerical rates in seconds, the oscillation frequency is $\Omega_m/\hbar$, and the phase is $\Omega_mt/\hbar$. The Debye benchmark explicitly restores SI constants. We identify $\theta_v=\theta_{12}$ and use $E$ and $E_\nu$ for the neutrino energy.

For two flavors $(\nu_e,\nu_x)$, after removing a term proportional to the identity, the flavor-basis Hamiltonian is
\begin{equation}
H_0(r)=\frac{1}{2}\left[(V-\Delta\cos2\theta_v)\sigma_z+\Delta\sin2\theta_v\,\sigma_x\right],
\label{eq:H0}
\end{equation}
where
\begin{equation}
\Delta=\frac{\Delta m_{21}^2}{2E_\nu},\qquad
V(r)=\sqrt{2}G_F \bar n_e(r).
\end{equation}
The matter gap and mixing angle are
\begin{align}
\Omega_m&=\sqrt{(\Delta\cos2\theta_v-V)^2+\Delta^2\sin^22\theta_v},\\
\sin2\theta_m&=\frac{\Delta\sin2\theta_v}{\Omega_m},\\
\cos2\theta_m&=\frac{\Delta\cos2\theta_v-V}{\Omega_m}.
\label{eq:thetamat}
\end{align}
We use $\theta_{12}=33.68^\circ$ and $\Delta m_{21}^2=7.49\times10^{-5}\,\mathrm{eV}^2$, consistent with NuFIT 6.0 \cite{NuFIT2024}.

\subsection{Matter fluctuations and noncommuting coupling}

Write $n_e=\bar n_e+\delta n_e$ and $\delta V=\sqrt{2}G_F\delta n_e$.  In the instantaneous matter basis,
\begin{equation}
H_{\rm fluc}(t)=\frac{\delta V(t)}{2}\left[\cos2\theta_m\,\sigma_z^{(m)}+\sin2\theta_m\,\sigma_x^{(m)}\right].
\label{eq:hfluc}
\end{equation}

A consistent basis convention, with $H_m=U_m^\dagger H_0U_m$, is
\begin{equation}
U_m=\begin{pmatrix}\cos\theta_m&\sin\theta_m\\-\sin\theta_m&\cos\theta_m\end{pmatrix},
\quad H_m=-\frac{\Omega_m}{2}\sigma_z^{(m)}.
\end{equation}
Then $U_m^\dagger\sigma_zU_m=\cos2\theta_m\sigma_z^{(m)}+\sin2\theta_m\sigma_x^{(m)}$, as used above. For a varying density, the transformed Hamiltonian also contains $-i\hbar U_m^\dagger\dot U_m$; neglecting this term is the adiabatic approximation. The instantaneous-basis construction itself does not require adiabaticity.
This equation is the key geometry of neutrino-specific detection. The longitudinal term produces phase noise and samples the low-frequency spectrum, whereas the transverse term drives transitions and samples the spectrum near $\Omega_m$. Unlike an engineered qubit, the relative coupling weights are not free parameters: they are fixed by $E_\nu$ and the mean density through MSW mixing.

The two-zone benchmark used below is listed in Table~\ref{tab:benchmark}. It is intentionally a minimal, illustrative discretization of the inner solar region rather than a digitized standard solar model profile. The first zone uses the central benchmark employed in the heuristic Debye-cell-scale estimate; the second is chosen at a lower density and temperature to produce a distinct MSW response. Quantitative solar inference must ultimately replace this table by $n_e(r)$ and $T(r)$ from a standard solar model.

\begin{table}[tbp]
\caption{Illustrative two-zone benchmark. The second row is a controlled theoretical benchmark, not an experimental fit or an exact SSM shell.}
\label{tab:benchmark}
\centering
\begin{tabular}{lcccc}
\hline\hline
Zone & $n_e$ [m$^{-3}$] & $T$ [K] & $L/\Rsun$ & $E_{\rm res}$ [MeV]\\
\hline
I & $6.05\times10^{31}$ & $1.55\times10^7$ & 0.08 & 1.88\\
II & $2.50\times10^{31}$ & $1.10\times10^7$ & 0.12 & 4.55\\
\hline\hline
\end{tabular}
\end{table}

\begin{figure}[tbp]
\includegraphics[width=0.75\columnwidth]{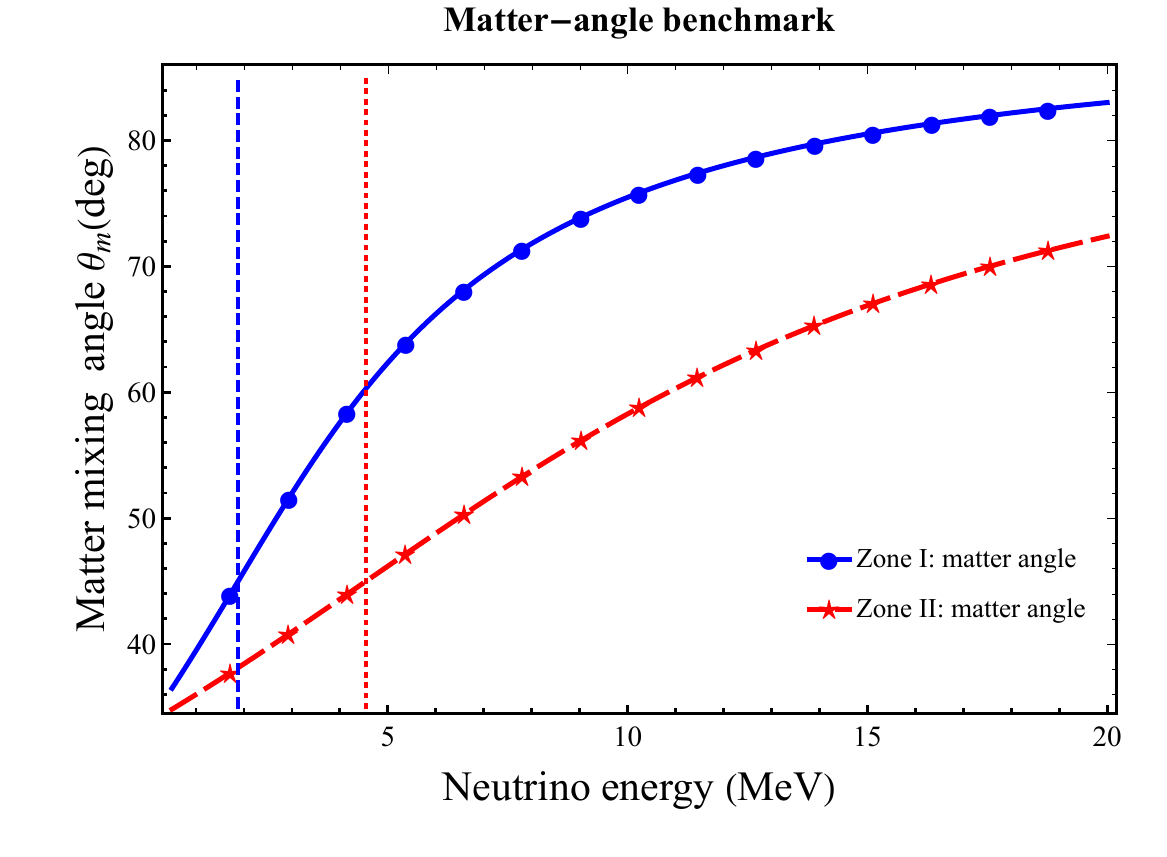}
\caption{Matter mixing angle for the two-zone benchmark. The dashed vertical lines mark the local MSW resonance energies. Distinct density-dependent rotations imply distinct environmental-coupling geometries at the same neutrino energy.}
\label{fig:mswangles}
\end{figure}

Figure~\ref{fig:mswangles} shows that the two densities produce distinct energy-dependent matter angles. The vertical resonance locations differ by more than a factor of two. This separation is precisely what later allows different energies to weigh the two zones differently.

Pure-dephasing thermometry also requires care in interpreting backaction: conservation of the probe's bare energy does not by itself establish thermodynamic noninvasiveness of the full probe--environment process \cite{Albarelli2023}. We use dephasing as a reduced-state encoding model, without making a noninvasiveness claim.

\begin{figure}[tbp]
\includegraphics[width=0.75\columnwidth]{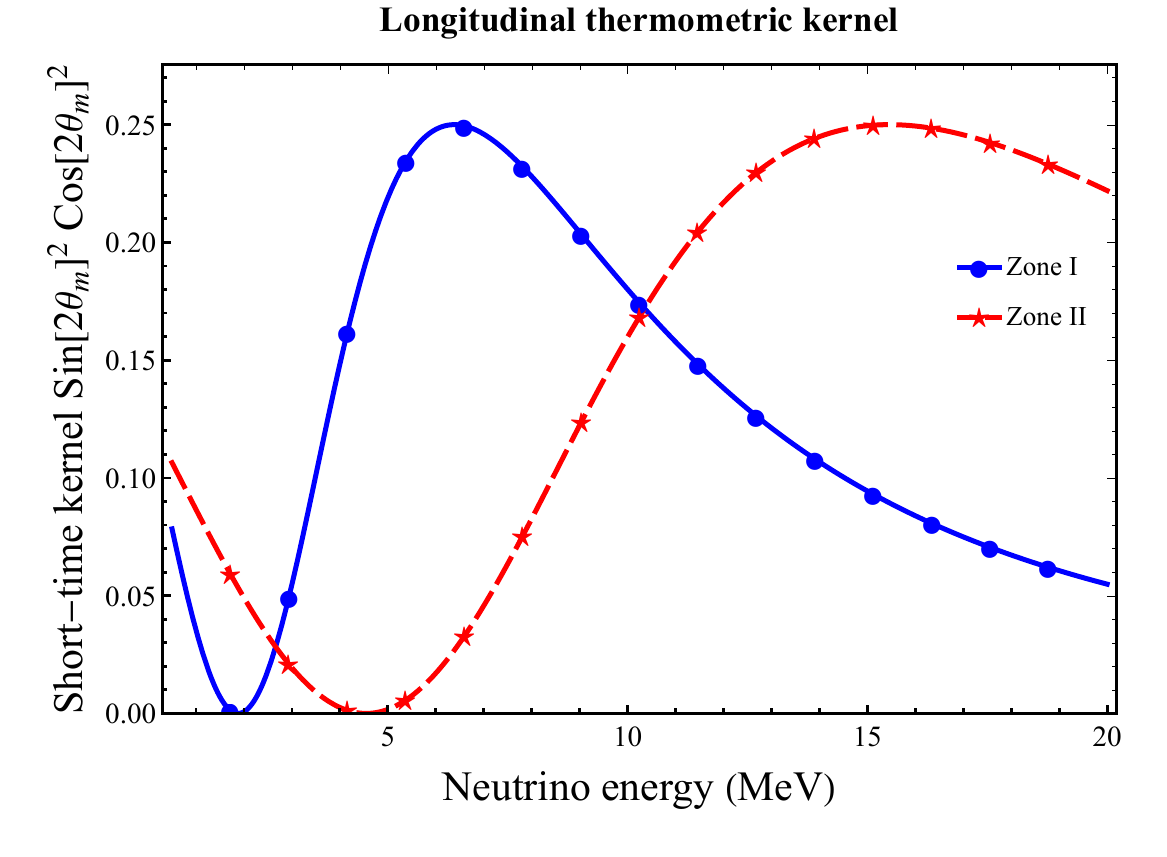}
\caption{Short-time longitudinal thermometric kernel $\sin^2(2\theta_m)\cos^2(2\theta_m)$ for the two benchmark densities.  Its maxima are displaced from the local MSW resonances, illustrating that maximal flavor conversion and maximal longitudinal thermal imprint are different conditions.}
\label{fig:kernels}
\end{figure}

For short-time longitudinal thermometry, the relevant MSW projection contains the product $\sin^22\theta_m\cos^22\theta_m$. Figure~\ref{fig:kernels} displays this kernel. The thermal-noise optimum is displaced from the MSW flavor-conversion resonance because the longitudinal contribution vanishes exactly where the transverse coupling is maximal; total decoherence need not vanish there. This difference between conversion matching and thermometric matching is an intrinsic feature of the neutrino implementation.

\section{Microscopic fluctuation spectrum}

For a stationary plasma, define
\begin{equation}
S_{nn}(\bm q,\omega)=\int d^3r\,dt\,e^{i(\omega t-\bm q\cdot\bm r)}\langle\delta n_e(\bm r,t)\delta n_e(0,0)\rangle.
\label{eq:snn}
\end{equation}
At equilibrium, the fluctuation--dissipation theorem relates this spectrum to the retarded density susceptibility \cite{Kubo1966,Ichimaru1992}, schematically
\begin{equation}
S_{nn}(q,\omega;T)=-\frac{2}{1-e^{-\omega/T}}\,\mathrm{Im}\,\chi^R_{nn}(q,\omega;T),
\label{eq:fdt}
\end{equation}
with $\chi^R_{nn}(\bm r,t)=-i\Theta(t)\langle[\delta n_e(\bm r,t),\delta n_e(0,0)]\rangle$ in the stated natural units and the same Fourier convention as Eq.~\eqref{eq:snn}. For angular frequency in SI, define the susceptibility with $1/\hbar$, replace the numerator by $2\hbar$, and replace $\omega/T$ by $\hbar\omega/(k_BT)$.

A propagating neutrino wave packet does not sample $S_{nn}$ at a single laboratory frequency. For a packet with form factor $F_{\rm wp}(\bm q)$ centered on $\bm r\simeq\bm vt$, the effective temporal spectrum is
\begin{equation}
S^{\rm traj}_{nn}(\Omega)=\int\frac{d^3q}{(2\pi)^3}|F_{\rm wp}(\bm q)|^2S_{nn}(\bm q,\Omega+\bm q\cdot\bm v),
\label{eq:traj}
\end{equation}
and the weak-potential spectrum is $S_V^{\rm traj}=2G_F^2S_{nn}^{\rm traj}$.  Equation~\eqref{eq:traj} is the appropriate object for a microscopic plasma calculation. A random Ohmic spectral density with a freely chosen coupling constant is not sufficient for a quantitative solar prediction.

In a weak-coupling secular description, the matter-basis density matrix obeys
\begin{equation}
\begin{aligned}
\dot\rho={}&-i[H_m+H_{\rm LS},\rho]+\frac{\gamma_\phi}{2}\Diss[\sigma_z]\rho\\
&+\gamma_\downarrow\Diss[A_\downarrow]\rho+\gamma_\uparrow\Diss[A_\uparrow]\rho.
\end{aligned}
\label{eq:master}
\end{equation}

We define $\Diss[A]\rho=A\rho A^\dagger-\{A^\dagger A,\rho\}/2$. Thus $\gamma_\phi$ is the longitudinal coherence-decay rate, not the coefficient multiplying $\Diss[\sigma_z]$. In the above energy ordering, $A_\downarrow=|1_m\rangle\langle2_m|$ and $A_\uparrow=A_\downarrow^\dagger$. For the unsymmetrized potential spectrum and the coupling in Eq.~\eqref{eq:hfluc}, the natural-unit secular rates are
\begin{align}
\gamma_\phi&=\tfrac12\cos^22\theta_m\,S_V^{\rm traj}(0),\\
\gamma_{\downarrow,\uparrow}&=\tfrac14\sin^22\theta_m\,S_V^{\rm traj}(\pm\Omega_m).
\end{align}
When $S_V$ uses an energy-valued potential and angular frequency in SI, divide these expressions by $\hbar^2$ and use $\pm\Omega_m/\hbar$. This fixes the proportionality conventions in the following equations, where
\begin{align}
\gamma_\phi&\propto \cos^22\theta_m\,S_V^{\rm traj}(0;T),\\
\gamma_\downarrow&\propto \sin^22\theta_m\,S_V^{\rm traj}(+\Omega_m;T),\\
\gamma_\uparrow&\propto \sin^22\theta_m\,S_V^{\rm traj}(-\Omega_m;T).
\label{eq:rates}
\end{align}
The stationary-medium equilibrium relation must be imposed before the trajectory integral. It does not generally imply $\gamma_\uparrow/\gamma_\downarrow=e^{-\Omega_m/T}$ for a moving probe. The Doppler-shifted frequency gives a Boltzmann factor depending on $\bm q\cdot\bm v$ inside the integral. Equation~\eqref{eq:rates} makes the physical meaning of the MSW rotation transparent: energy changes both the coupling direction and the spectral frequency probed by the neutrino.

A quantitative microscopic solar calculation should evaluate $\chi^R_{nn}$ in random-phase approximation or kinetic theory, including screening, ions, collisional effects, and the trajectory form factor. Here, the electron-only Debye-cell construction is used only as a heuristic absolute-scale benchmark after the information-theoretic structure has been established; it is not a controlled evaluation of $S_{nn}^{\rm traj}$ or a rigorous upper bound.

\section{Two-zone thermometry and conditional single-energy rank deficiency}

Consider two effective regions with temperatures $(T_1,T_2)$ and fixed mean densities $(n_{e1},n_{e2})$. The additive construction below is a restricted longitudinal model. We assume adiabatic transport of the same instantaneous matter-eigenstate labels across the two effective regions, negligible transverse transitions, statistically independent zone noise, and temperature-independent preparation, mean Hamiltonian, and Lamb-shift phase. The constant-zone entries in Table~\ref{tab:benchmark} are therefore integration benchmarks rather than a literal sequence of two abrupt slabs; an abrupt interface would require explicit basis matching and need not reduce to one scalar exponent. Energies near an MSW resonance do not by themselves justify the neglect of transverse noise.\par

In the longitudinal regime, the accumulated dephasing exponent is
\begin{equation}
\Gamma(E;T_1,T_2)=\Gamma_1(E,T_1)+\Gamma_2(E,T_2),
\label{eq:gammatotal}
\end{equation}
with
\begin{equation}
\Gamma_j(E,T_j)=\int_{\mathcal R_j}\frac{dr}{c}\,\gamma_{\phi,j}[E,n_{ej}(r),T_j].
\end{equation}
For the constant-zone benchmark, this reduces to $\Gamma_j=\gamma_{\phi,j}L_j/c$.

To isolate the identifiability structure, introduce logarithmic thermal parameters
\begin{equation}
\vartheta_j=\ln(T_j/T_{j,0}).
\end{equation}
If the temperature dependence of the reduced state enters only through the scalar $\Gamma$, then $\partial_{\vartheta_j}\rho=(\partial_\Gamma\rho)(\partial_{\vartheta_j}\Gamma)$.  The QFIM must therefore factorize as
\begin{equation}
\QFI_{ij}^{(E)}=K(E,\Gamma)\,g_i(E)g_j(E),\qquad
g_i(E)=\partial_{\vartheta_i}\Gamma,
\label{eq:outer}
\end{equation}
for some non-negative scalar $K$.  For an equatorial dephasing qubit, $K=(e^{2\Gamma}-1)^{-1}$.  For an electron-flavor state, the prefactor is modified by a positive MSW projection factor, but the outer-product structure is unchanged as long as the temperatures affect only $\Gamma$.

Equation~\eqref{eq:outer} immediately gives
\begin{equation}
\boxed{\det\QFI^{(E)}=0.}
\label{eq:rankone}
\end{equation}
This is a structural statement: a single-energy scalar dephasing channel cannot identify two independent temperature parameters, regardless of the number of detected neutrinos(see appendix~\ref{app:rankone} for the full derivation).  Increasing statistics shrinks uncertainty only along the one thermal combination to which the channel is sensitive; the orthogonal combination remains locally unconstrained.

Under the Debye scaling used below, $\gamma_\phi\propto T^{-1}$ at fixed density, so $g_j=-\Gamma_j$ for $\vartheta_j=\ln T_j$.  The direction of the thermal-sensitivity vector is therefore controlled by the relative accumulated exponents in the two zones.  Because the dephasing weights contain $\cos^22\theta_{mj}(E)$, this direction is energy dependent.

\section{Energy diversity restores multiparameter identifiability}
\label{sec:energy}

Let independent event classes probe energies $E_1$ and $E_2$. Fisher information adds for independent samples,
\begin{equation}
\QFI_{\rm tot}=M_1\QFI^{(E_1)}+M_2\QFI^{(E_2)}.
\end{equation}
Writing $\QFI^{(E_a)}=K_a\bm g_a\bm g_a^T$, one finds for two equally parameterized temperatures
\begin{equation}
\det\QFI_{\rm tot}=M_1M_2K_1K_2\left(g_{11}g_{22}-g_{12}g_{21}\right)^2.
\label{eq:dettwo}
\end{equation}
Hence
\begin{equation}
\det\QFI_{\rm tot}>0\quad\Longleftrightarrow\quad \bm g(E_1)\not\parallel\bm g(E_2),
\label{eq:criterion}
\end{equation}
provided both channels carry nonzero information.

Both event counts must also be positive, and the energy labels must be retained. The simple sum assumes parameter-independent sampling weights; unobserved energy mixing or a temperature-dependent source spectrum defines a different statistical model. The result is geometrically simple: distinct energies are useful only if they interrogate the thermal profile in distinct directions.

To visualize the geometry independently of the tiny absolute solar scale, we define the normalized identifiability index
\begin{equation}
\ident=\frac{4\det\QFI}{(\Tr\QFI)^2},\qquad 0\leq\ident\leq1.
\label{eq:ident}
\end{equation}
For a positive $2\times2$ matrix, $\ident=0$ signals rank deficiency and $\ident=1$ corresponds to equal Fisher eigenvalues.

This index is defined only for $\Tr\QFI>0$ and for the chosen logarithmic-temperature coordinates and sample allocation. It is invariant under one common positive scale factor, but not under arbitrary coordinate changes or separate rescalings of energy channels.

\begin{figure}[tbp]
\includegraphics[width=0.75\columnwidth]{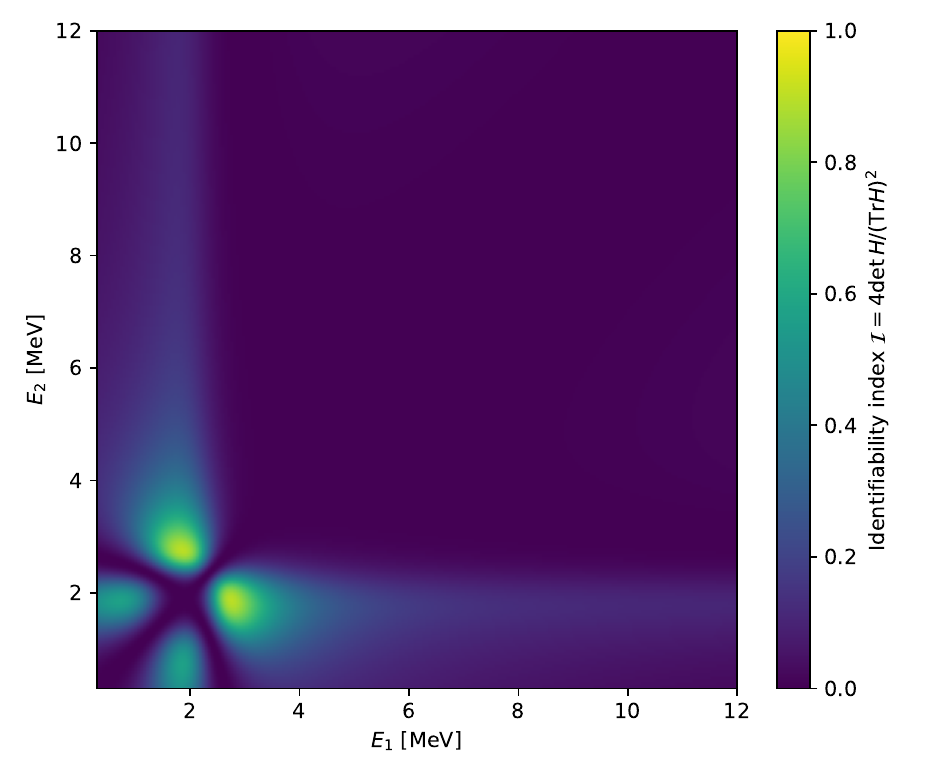}
\caption{Normalized two-temperature identifiability index $\ident=4\det\QFI/(\Tr\QFI)^2$ for two neutrino energies.  The diagonal is rank deficient. Bright regions occur when MSW energy dependence produces strongly noncollinear thermal-sensitivity vectors.}
\label{fig:identmap}
\end{figure}

Figure~\ref{fig:identmap} shows $\ident(E_1,E_2)$ for the benchmark in Table~\ref{tab:benchmark}.  The diagonal is dark because identical energies give parallel sensitivity vectors. High identifiability appears when one energy is close to the sharp MSW rotation of Zone~I, while the second energy lies where the relative weighting of Zone~II is substantially different. A representative pair $(1.9,2.7)$ MeV gives $\ident\simeq0.89$, whereas $(6,8)$ MeV gives only $\ident\simeq1.3\times10^{-3}$.  The corresponding QFIM condition numbers are approximately 2 and $3.0\times10^3$, respectively. These numerical values depend on the benchmark, but the rank criterion in Eq.~\eqref{eq:criterion} does not. We stress that the best-conditioned pair lies near or below the analysis thresholds of current $^8$B measurements, for example, 3~MeV of recoil-electron kinetic energy in Borexino~\cite{Borexino2020}. A 1.9~MeV neutrino gives recoil electrons of at most $\simeq1.7$~MeV. In elastic $\nu$--$e$ scattering, the neutrino energy is moreover not reconstructed event by event, so the ``energy labels'' assumed in Eq.~\eqref{eq:criterion} are smeared by the recoil kinematics. Both effects reduce the noncollinearity actually available to a detector, and they must be included through the event-level Fisher information discussed in Sec.~\ref{sec:flavor}.\par

\begin{figure}[tbp]
\includegraphics[width=0.75\columnwidth]{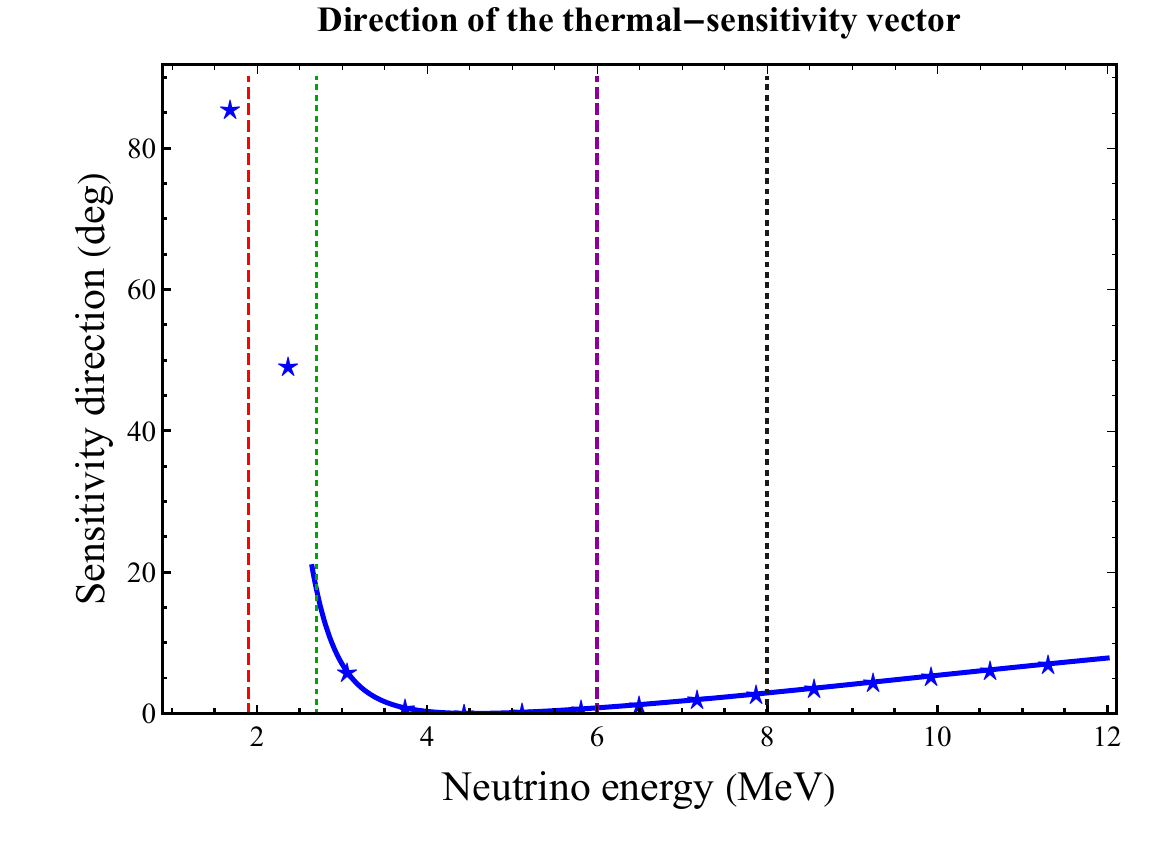}
\caption{Direction of the logarithmic thermal-sensitivity vector $\bm g(E)$.  Rapid rotations occur in the MSW-sensitive energy region. The marked pairs (vertical lines and open circles at 1.9, 2.7, 6, and 8 MeV; $\alpha\simeq89.9^\circ$, $17.5^\circ$, $0.8^\circ$, and $2.9^\circ$) illustrate why $(1.9,2.7)$ MeV produces a well-conditioned two-parameter geometry, whereas $(6,8)$ MeV does not.}
\label{fig:direction}
\end{figure}

\begin{figure}[tbp]
\includegraphics[width=0.75\columnwidth]{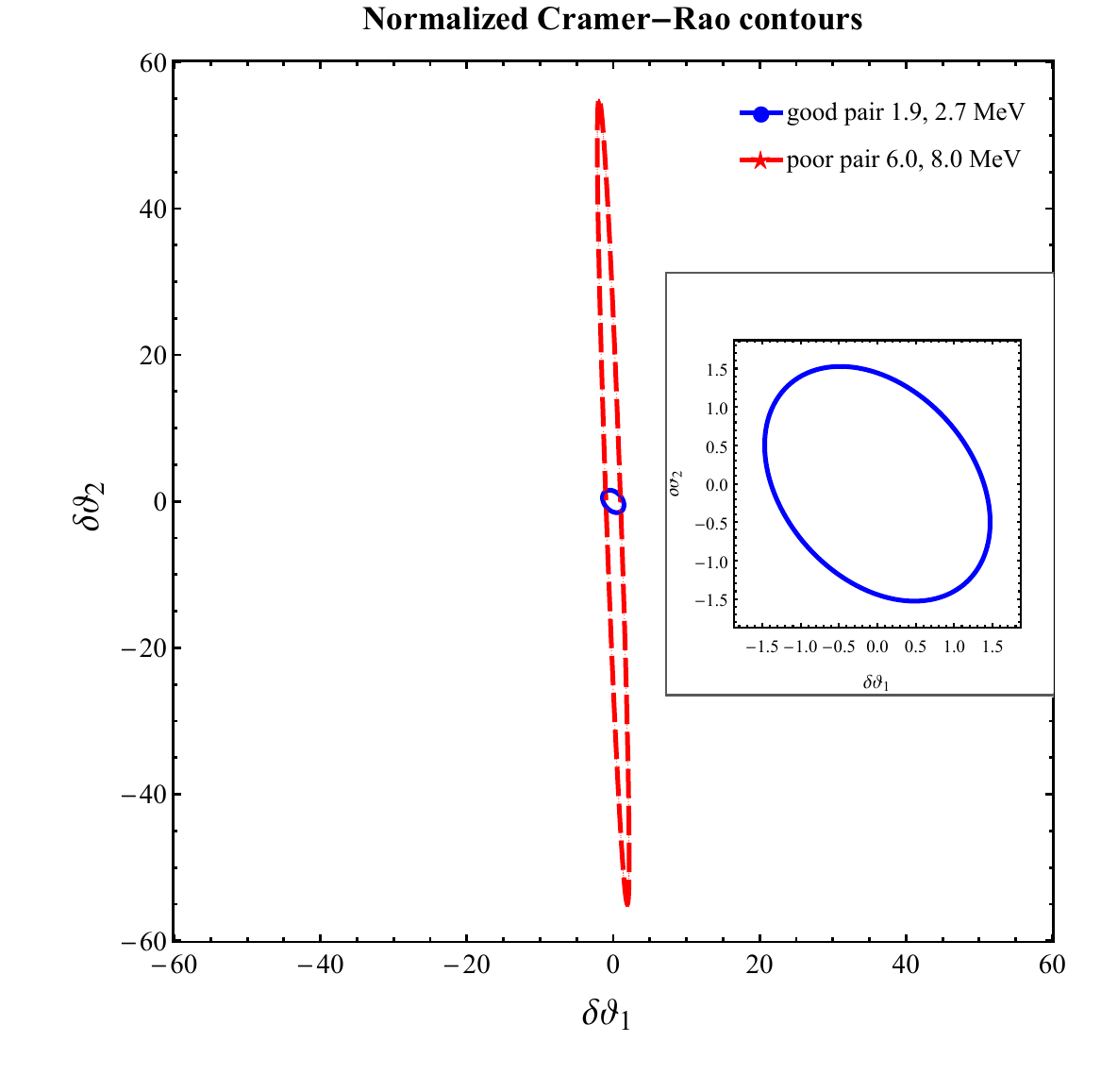}
\caption{Normalized Cram\'er--Rao geometry in the logarithmic-temperature coordinates $\delta\vartheta_j$. For each energy pair we define $\bar{\mathcal H}=\mathcal H_{\rm tot}/\mathrm{Tr}\,\mathcal H_{\rm tot}$ and plot the unit quadratic contour $\delta\bm\vartheta^{T}\bar{\mathcal H}\,\delta\bm\vartheta=1$. The complete extent of both contours is shown. This normalization removes the overall Fisher scale, so the curves display conditioning rather than an experimental confidence region.}
\label{fig:ellipses}
\end{figure}

For Figs.~\ref{fig:identmap}--\ref{fig:ellipses} we use equatorial preparations, equal event counts in the two labeled energy classes, $\Gamma_j=\gamma_{0j}\cos^2(2\theta_{mj})L_j/c$, and the fixed-density Debye scaling $\gamma_{0j}\propto n_{ej}^2/T_j$. Because the physical Debye exponents are $\ll1$, the plotted normalized geometry is evaluated in the small-$\Gamma$ limit $K\simeq(2\Gamma)^{-1}$; the unknown common proportionality constant then cancels. The companion  source code (see Data Availability) reproduces $\mathcal I=0.89070$ and $\kappa=1.988$ for $(1.9,2.7)$ MeV, and $\mathcal I=1.3204\times10^{-3}$ and $\kappa=3.027\times10^3$ for $(6,8)$ MeV. These are benchmark-dependent shape diagnostics, not absolute errors.

Figure~\ref{fig:direction} displays the sensitivity direction
\begin{equation}
\alpha(E)=\tan^{-1}\left(\frac{|g_2(E)|}{|g_1(E)|}\right).
\end{equation}
Around 1.9 MeV, the first-zone longitudinal weight nearly vanishes at its local resonance, so the sensitivity vector points almost entirely toward the second thermal parameter. By 2.7 MeV, the vector has rotated strongly. At 6--8 MeV, however, the direction changes little, explaining the near singularity of the combined high-energy QFIM.

The same conclusion is visible in Cram\'er--Rao geometry. Figure~\ref{fig:ellipses} compares quadratic confidence contours after normalizing away the overall Fisher scale. The good energy pair yields a compact ellipse, whereas the poor pair remains extremely elongated. This plot should not be interpreted as an absolute experimental error ellipse; it isolates the shape of the inverse problem. The absolute solar errors are addressed below.

The physical message is broader than the two-zone construction. If a continuous thermal profile is expanded as $T(r)=T_{\rm ref}(r)[1+\sum_k a_k B_k(r)]$, then each energy defines a sensitivity vector in the coefficient space $(a_1,a_2,\ldots)$. Energy diversity increases the rank only to the extent that these vectors span different directions. Thus, solar neutrino thermometry is naturally an inverse problem whose reconstructible thermal modes are determined by the singular-value spectrum of the Fisher operator.

\section{Two probes, local baths, and spatial correlations}

We now distinguish a genuine two-probe resource from the two-mode representation of one neutrino. Let probes $A$ and $B$ experience Gaussian phase noise with local rates and cross correlations. A standard correlated-dephasing generator has the form
\begin{equation}
\dot\rho=-i[H_S,\rho]+\sum_{\alpha,\beta=A,B}\gamma_{\alpha\beta}\left[\sigma_z^{(\alpha)}\rho\,\sigma_z^{(\beta)}-\tfrac12\{\sigma_z^{(\beta)}\sigma_z^{(\alpha)},\rho\}\right].
\label{eq:twoprobeME}
\end{equation}

For the symmetric case, set $\gamma_{AA}=\gamma_{BB}=\gamma_\phi/2$, so that the single-probe coherence exponent is $u=2\gamma_{AA}t=\gamma_\phi t$. Complete positivity requires the matrix $(\gamma_{\alpha\beta})$ to be positive semidefinite. Define $|\Phi^\pm\rangle=(|00\rangle\pm|11\rangle)/\sqrt2$ and $|\Psi^\pm\rangle=(|01\rangle\pm|10\rangle)/\sqrt2$.
We write the normalized spatial correlation as
\begin{equation}
\mu(d)=\exp(-d/\ell_c),\qquad
\gamma_{AB}=\mu\sqrt{\gamma_{AA}\gamma_{BB}}.
\label{eq:mu}
\end{equation}
The local-bath limit is $d\gg\ell_c$, while a common bath corresponds to $d\ll\ell_c$.

For symmetric Gaussian dephasing, Bell coherences acquire sum and difference phases. In scaled variables, their exponents may be written
\begin{equation}
\Gamma_\Phi=2u(1+\mu),\qquad
\Gamma_\Psi=2u(1-\mu).
\label{eq:bellgamma}
\end{equation}
Here $u$ is the single-probe dephasing exponent defined above. Thus $\Phi$ sectors amplify common-mode fluctuations, whereas $\Psi$ sectors become decoherence-free when the noise is perfectly common and become sensitive as the trajectories decorrelate. Below we write $\Gamma_+\equiv\Gamma_\Phi$ and $\Gamma_-\equiv\Gamma_\Psi$.

For independent coordinates $(\ln T,\ln\ell_c)$, fixed $d$, $x=d/\ell_c$, and $u\propto T^{-1}$,
\begin{equation}
\bm v_\pm=\begin{pmatrix}-2u(1\pm\mu)\\\pm2ux\mu\end{pmatrix},\qquad
\QFI_\pm=\frac{\bm v_\pm\bm v_\pm^T}{e^{2\Gamma_\pm}-1}.
\end{equation}
A single sector is rank one. Independent samples of both sectors instead give
\begin{equation}
\det\QFI_{\rm pair}=\frac{M_+M_-(8u^2x\mu)^2}{(e^{2\Gamma_+}-1)(e^{2\Gamma_-}-1)},
\end{equation}
which is positive for nonzero $u$, both positive counts, and finite nonzero $x$. The diagonal curves alone are conditional sensitivities, not joint estimation errors. If a microscopic model relates $\ell_c$ to $T$, its chain rule must be included.

\begin{figure}[tbp]
\includegraphics[width=0.75\columnwidth]{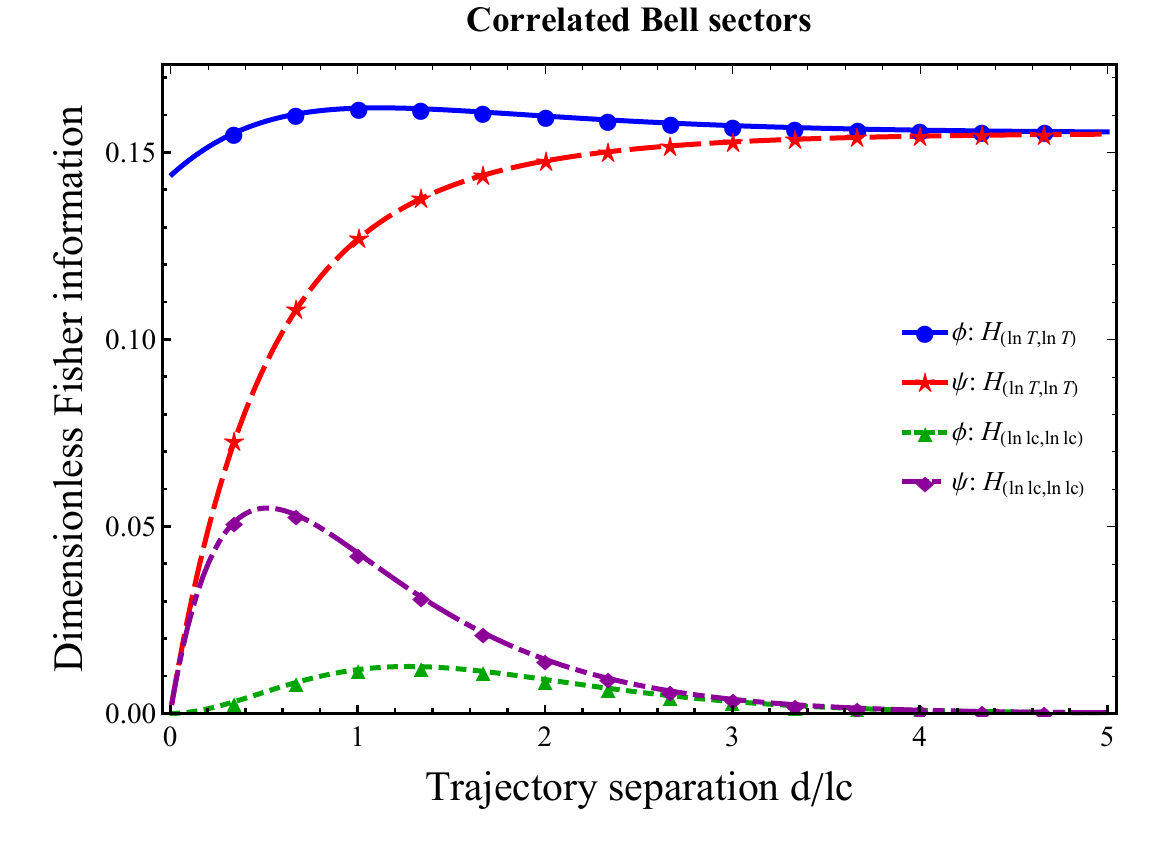}
\caption{Scaled Fisher information of correlated Bell sectors versus $x=d/\ell_c$ for $u=0.3$ in the coordinates $(\ln T,\ln\ell_c)$ at fixed $d$. The plotted diagonal elements are $\QFI_{\ln T,\ln T}$ and $\QFI_{\ln\ell_c,\ln\ell_c}$ for the $\Phi$ and $\Psi$ sectors. The exponential law $\mu=e^{-d/\ell_c}$ is a phenomenological correlation model, not a plasma-derived result.}
\label{fig:bell}
\end{figure}

Figure~\ref{fig:bell} illustrates the complementary information channels. At fixed $u$, the $\Phi$ thermal information is $q(2u(1+\mu))$, where $q(z)=z^2/(e^{2z}-1)$, so its maximum can occur at intermediate correlation. The plotted $\Psi$ response grows as the paths decorrelate for the illustrated regime; neither monotonicity is universal in $u$. Sensitivity to $\ell_c$ is largest at intermediate separation, because both $d\mu/d\ln\ell_c=(d/\ell_c)\mu$ and a nontrivial dephasing response are required. Consequently, a two-probe experiment could, in principle, use different sectors or separations to separate temperature from spatial correlation length. In an astrophysical solar setting, however, prepared Bell pairs of neutrinos are not a realistic resource; this calculation is an information-theoretic benchmark and a guide for quantum-simulator implementations.

For the plotted value $u=0.3$, the $\Phi$ thermal-information curve reaches its maximum at $d/\ell_c\simeq1.115$; thus, common-mode sensitivity is enhanced at small separation, but its maximum is not at $d=0$ for this choice of $u$.

Correlations can also encode thermometric information in collisional protocols, in which sequential ancillas interact with an intermediate system \cite{Seah2019}. That architecture differs from independent solar-neutrino events and from the simultaneous two-probe model above; its advantages cannot be assigned to the present setup without specifying the corresponding interactions and resource budget.

\begin{figure}[tbp]
\includegraphics[width=0.75\columnwidth]{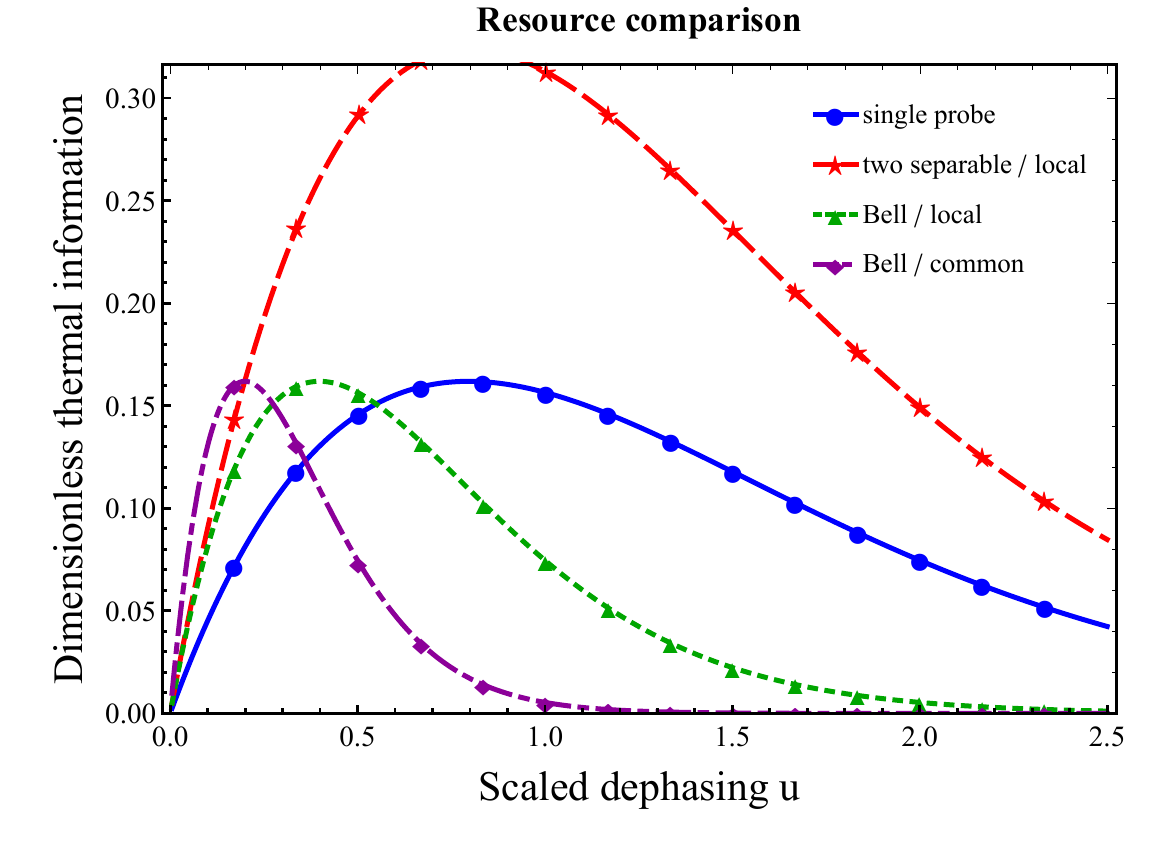}
\caption{Dimensionless thermal information in a scaled dephasing benchmark.  Two local probes give the additive reference. A common-bath Bell sector can dominate at sufficiently short times, but this redistribution of information does not modify the microscopic weak-interaction scale.}
\label{fig:resources}
\end{figure}

To separate a genuine correlation advantage from resource counting, Fig.~\ref{fig:resources} compares a single probe, two independent local probes, a Bell pair in local baths, and a Bell sector in a common bath for a simple dephasing model. At very short times, a common-bath Bell coherence can write information more rapidly, while at longer times the two independent probes retain a broader useful window. This reproduces the qualitative hierarchy known from two-qubit thermometry \cite{Gebbia2020}: entanglement may be favorable under a time constraint, whereas independent coherent probes provide the robust additive benchmark when interrogation time is not fixed.

For $u\propto T^{-1}$, the four conditional per-trial QSNRs are $q(u)$, $2q(u)$, $q(2u)$, and $q(4u)$, respectively; a two-probe trial consumes two neutrinos. Comparisons at a fixed event number, a fixed particle number, and a fixed total interrogation time must be distinguished.

\section{Encoded information versus flavor-accessible information}
\revision{\label{sec:flavor}}

\begin{figure}[tbp]
\includegraphics[width=0.75\columnwidth]{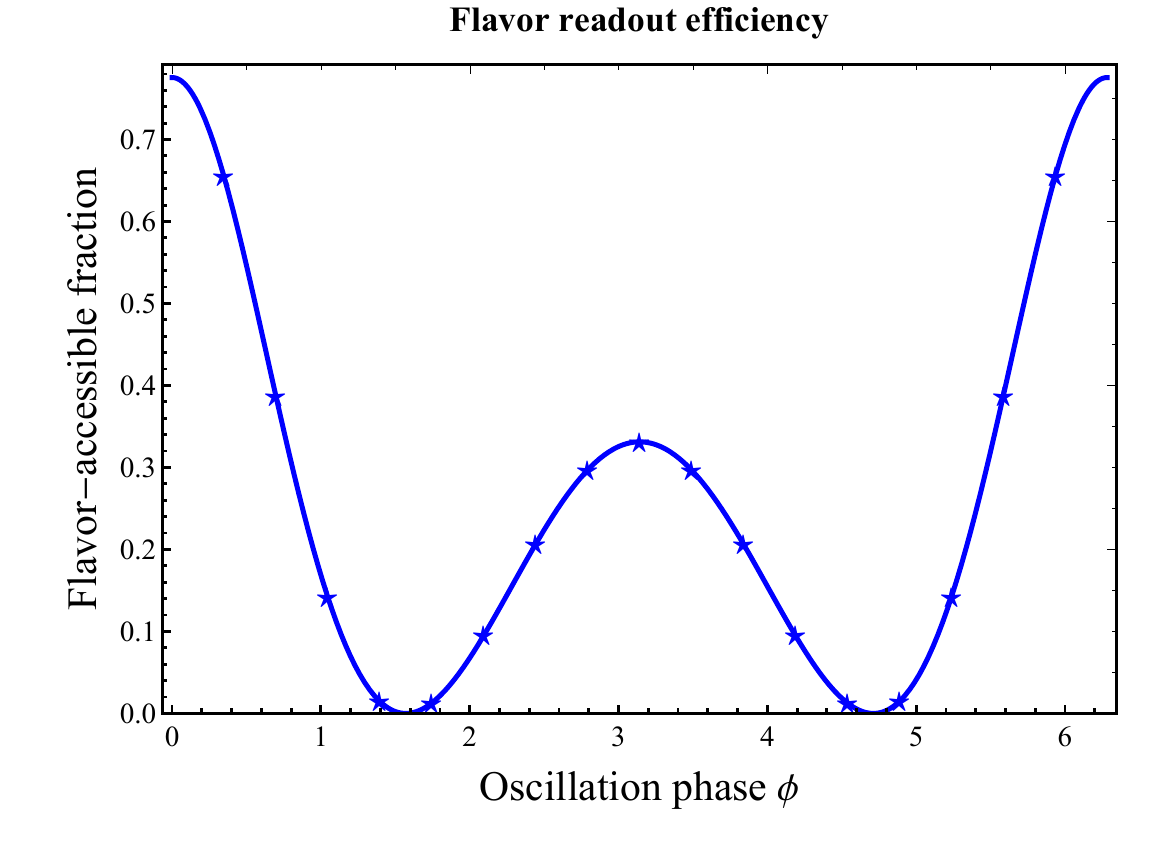}
\caption{Fraction of the temperature QFI accessible to ideal flavor projection at $u_\star=0.796812\ldots$ using the Zone-I matter angle at 10 MeV, $\theta_m=75.482^\circ$. Flavor readout can be efficient at favorable phases and is blind at oscillation quadrature.}
\label{fig:flavorfraction}
\end{figure}

The QFI assumes access to the optimal quantum measurement. For solar neutrinos, the operational measurement is primarily flavor dependent. For a coherent electron-flavor state prepared and measured in the same constant-density longitudinal model, the survival probability is
\begin{equation}
P_{ee}=1-\frac{\sin^22\theta_m}{2}\left[1-e^{-u}\cos\phi\right],
\label{eq:pee}
\end{equation}
where $u=\gamma_\phi t$ and $\phi=\Omega_m t$ in natural units ($\phi=\Omega_m t/\hbar$ for SI time).  For $u\propto T^{-1}$,
\begin{equation}
T^2\FI_T^{\rm flav}=\frac{\sin^42\theta_m\,e^{-2u}u^2\cos^2\phi}{4P_{ee}(1-P_{ee})},
\label{eq:flavfi}
\end{equation}
whereas the total-state QSNR is
\begin{equation}
Q_T=\sin^22\theta_m\frac{u^2}{e^{2u}-1}.
\label{eq:qt}
\end{equation}
The ratio is phase-dependent. Figure~\ref{fig:flavorfraction} shows that flavor projection may recover a large fraction of the QFI at favorable phases but becomes blind at oscillation quadrature. This difference is consistent with the broader lesson of recent neutrino-estimation studies: a state can encode information that is inaccessible to the experimentally available readout \cite{Frugiuele2026,Huang2026}.

Finite measurement resolution can alter attainable thermometric precision. This issue is explicit in low-temperature bounds \cite{Potts2019,Jorgensen2020}; those asymptotic bounds are not solar-temperature limits, but reinforce the need to distinguish state QFI from information retained by a specified detector. For a realistic solar analysis, Eq.~\eqref{eq:flavfi} must be replaced by event-level Fisher information obtained after folding the oscillation probability with the production spectrum, cross sections, detector efficiency, and energy resolution. Source averaging is essential because solar neutrinos are produced over an extended region, and kinematical wave-packet separation must be distinguished from thermal decoherence.

In fact, complete unobserved phase averaging in this restricted model yields $\overline P_{ee}=1-\sin^22\theta_m/2$, independent of $u$. Hence, the flavor FI for temperature vanishes. The averaged state also has temperature-independent populations and zero thermal QFI when only the longitudinal damping depends on temperature. Information must be evaluated after averaging the state or event probabilities, unless the averaging variable is experimentally recorded. Surviving sensitivity from transverse transitions, mean refraction, or production rates requires a separate calculation.

\section{Absolute solar scale: heuristic Debye-cell benchmark}
\label{sec:debye}

The previous sections addressed information geometry. We now restore the absolute solar scale. For a weakly coupled ionized plasma, the Debye length and particle number in a Debye sphere are
\begin{equation}
\lambda_D=\left(\frac{\epsilon_0k_BT}{n_e e^2}\right)^{1/2},\qquad
N_D=\frac{4\pi}{3}n_e\lambda_D^3.
\label{eq:debye}
\end{equation}
A heuristic local number-fluctuation estimate takes $\sigma_n/n_e\sim N_D^{-1/2}$ and a crossing correlation time $\tau_D\sim\lambda_D/c$.  With $\sigma_V=V_0\sigma_n/n_e$, the unprojected scale is
\begin{equation}
\gamma_0\sim\frac{\sigma_V^2\tau_D}{\hbar^2},\qquad
\gamma_\phi=\cos^22\theta_m\,\gamma_0.
\label{eq:gammaDebye}
\end{equation}

Here $V_0=\sqrt2G_Fn_e$ is the uniform mean potential. For classical exponential potential correlations, the Gaussian cumulant gives
\begin{equation}
u(t)=\frac{\cos^22\theta_m\sigma_V^2}{\hbar^2}
\left[\tau_Dt-\tau_D^2(1-e^{-t/\tau_D})\right].
\end{equation}
For $t\gg\tau_D$ this gives $\gamma_\phi=\cos^22\theta_m\,\sigma_V^2\tau_D/\hbar^2$, consistent with the secular rate $\gamma_\phi=\tfrac12\cos^22\theta_m S_V(0)/\hbar^2$ for the exponential spectrum $S_V(0)=2\sigma_V^2\tau_D$. The linear exponent used here applies for $t\gg\tau_D$; $u\ll1$ does not mean $t\ll\tau_D$. The estimate also implies $\gamma_0\propto n_e^2/T$ and uses an electron-only screening length.

For the central parameters, Eqs.~\eqref{eq:debye} give $\lambda_D=3.493\times10^{-11}\,\mathrm{m}$, $N_D=10.80$, and $\tau_D=\lambda_D/c=1.165\times10^{-19}\,\mathrm{s}$. The estimate assumes electron-only Debye screening, independent Poisson-like cell-number fluctuations $\sigma_n/n_e\simeq N_D^{-1/2}$, and a pointlike trajectory on the cell scale. It omits ionic screening, collisional corrections, and the explicit wave-packet form factor in Eq.~\eqref{eq:traj}. Consequently, the resulting rate is a model-defined reference scale, not a controlled microscopic plasma prediction or an established upper bound. In particular, for a pointlike trajectory the variance of the line-integrated density depends on the transverse coarse-graining scale. Here that scale is fixed by hand at $\lambda_D$, whereas in Eq.~\eqref{eq:traj} it is set by the wave-packet form factor $F_{\rm wp}$. Density structure on scales shorter than the transverse coherence of the packet does not contribute a forward-scattering phase. It corresponds instead to incoherent scattering, which is discussed next.
At fixed $n_e$, this construction gives
\begin{equation}
\lambda_D\propto T^{1/2},\quad N_D\propto T^{3/2},\quad \sigma_V^2\propto T^{-3/2},\quad \tau_D\propto T^{1/2},
\end{equation}
and therefore
\begin{equation}
\gamma_0\propto T^{-1}.
\label{eq:Tscaling}
\end{equation}
This scaling is model-specific and should not be interpreted as a universal law of neutrino thermometry.

For the central benchmark $T_c=1.55\times10^7$ K and $n_e=6.05\times10^{31}\,\mathrm{m}^{-3}$, the heuristic estimate gives $\gamma_0\simeq1.46\times10^{-12}\,\mathrm{s}^{-1}$.  At 10 MeV, $\theta_m\simeq75.5^\circ$ and $\gamma_\phi\simeq1.1\times10^{-12}\,\mathrm{s}^{-1}$.  The positive fixed-event-count optimum of Eq.~\eqref{eq:qt}, at fixed mixing angle and for $u\propto1/T$, satisfies
\begin{equation}
u_\star=1-e^{-2u_\star},\qquad u_\star=0.796812130\ldots,
\end{equation}
so the optimal interaction time is $t_{\rm opt}=u_\star/\gamma_\phi$, of order $7\times10^{11}$ s.  The corresponding length is of order $3\times10^{11}\Rsun$. A solar transit is therefore deep in the $u\ll1$ regime.

\begin{figure}[tbp]
\includegraphics[width=0.75\columnwidth]{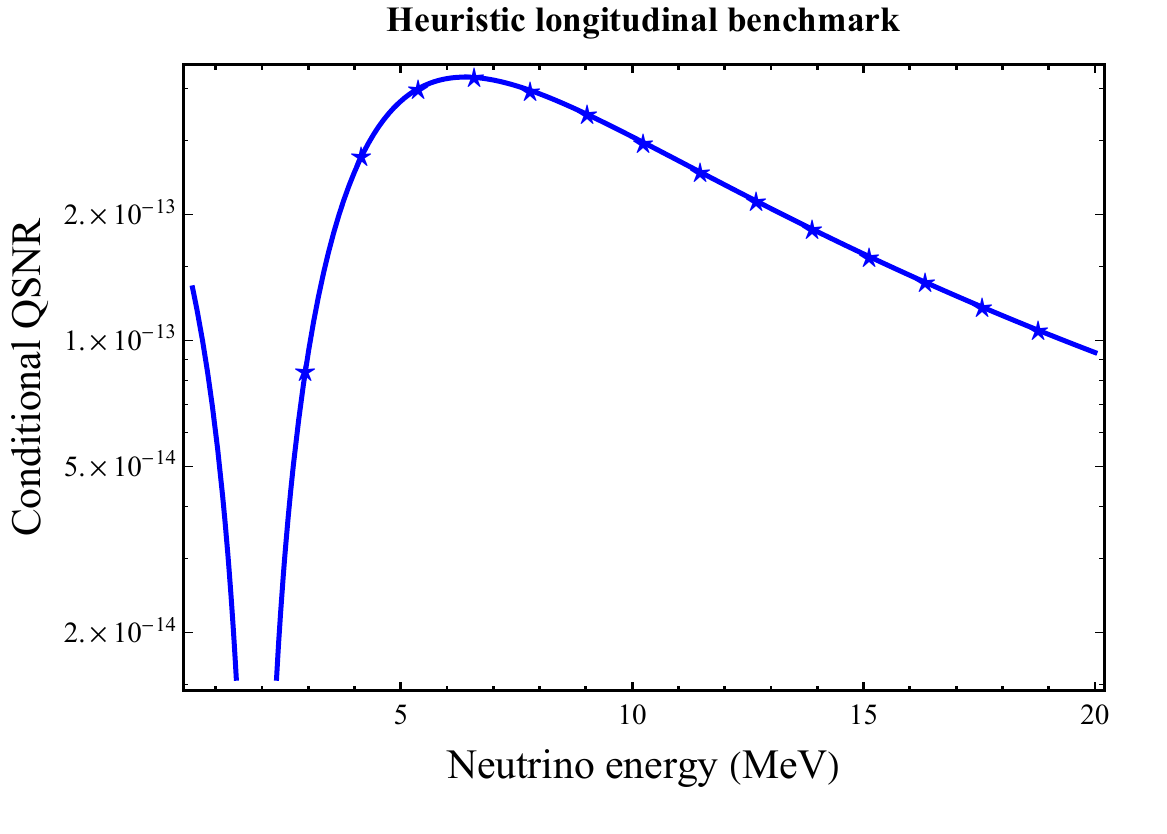}
\caption{Conditional coherent-state QSNR accumulated over one solar radius crossing in the heuristic electron-only Debye-cell model. The curve includes only the longitudinal channel and therefore does not represent full-channel solar sensitivity.}
\label{fig:solarqsnr}
\end{figure}

Figure~\ref{fig:solarqsnr} shows a conditional coherent-state QSNR for a uniform central-density, central-temperature medium over $R_\odot=6.957\times10^8$ m. This is distinct from the two-zone path of total length $0.20R_\odot$ and is not a propagation calculation through a solar profile. The dip at the local MSW resonance occurs because the longitudinal coupling vanishes there. Away from resonance, the maximum remains on the order of $10^{-13}$; the curve peaks at $Q_T\simeq4.2\times10^{-13}$ near 6.4 MeV, where the Zone-I longitudinal kernel of Fig.~\ref{fig:kernels} is maximal. At 10 MeV, we obtain
\begin{equation}
Q_T(\Rsun)\simeq3.06\times10^{-13}.
\label{eq:qt10}
\end{equation}
This reproduces the scale of the chosen longitudinal Debye-cell benchmark. It is not a fundamental bound on all temperature-dependent neutrino channels. Because solar $^8$B neutrinos reach the detector as an effectively incoherent mixture of matter (or mass) eigenstates~\cite{ChundawatLi2026}, Eq.~\eqref{eq:qt10} should be read as the information that a hypothetical coherent state would carry. The operational value in this purely longitudinal channel is zero (Sec.~\ref{sec:flavor}).

\paragraph{Comparison with collisional damping.} The Debye-cell exponent should be compared with the incoherent (non-forward) neutrino--electron scattering rate, which enters quantum kinetic equations as a collisional damping term that also destroys flavor coherence~\cite{SiglRaffelt1993,Tututi2002}. Using $\sigma_{\nu_e e}\sim10^{-43}\,\mathrm{cm}^2$ at 10 MeV, the central-density rate is $n_e\sigma_{\nu_e e}c\sim2\times10^{-7}\,\mathrm{s}^{-1}$, roughly five orders of magnitude above $\gamma_0$. Its temperature dependence, however, arises only through the thermal motion of the electrons and is therefore suppressed by powers of $k_BT/m_ec^2\simeq2.6\times10^{-3}$. A collisional thermometric QSNR therefore need not be smaller than the benchmark of Eq.~\eqref{eq:qt10}. We do not evaluate it here, but it must be bounded before the Debye-cell value can be regarded as representative of propagation-induced thermal information.

\begin{figure}[tbp]
\includegraphics[width=0.75\columnwidth]{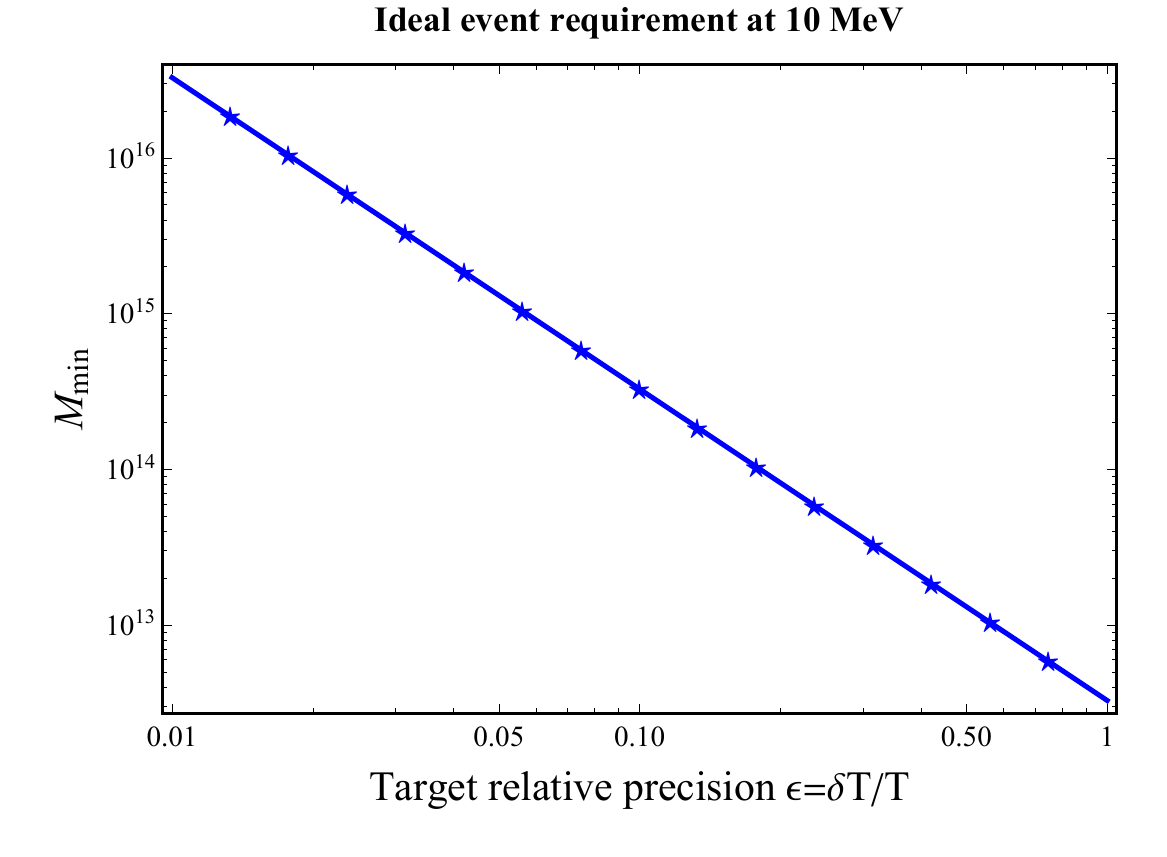}
\caption{Ideal local quantum-limit event requirement for a direct temperature estimate at 10 MeV in the one-solar-radius heuristic longitudinal benchmark. The count assumes a coherent state, known nuisance parameters, and access to the QFI; detector inefficiency, averaging, and nuisance profiling can only increase it.}
\label{fig:events}
\end{figure}

For a target relative precision $\epsilon=\delta T/T$, the ideal single-parameter event requirement is
\begin{equation}
M_{\min}=\frac{1}{\epsilon^2Q_T}.
\label{eq:mmin}
\end{equation}
Figure~\ref{fig:events} shows this requirement at 10 MeV. A 10\% estimate already requires approximately $3.3\times10^{14}$ ideal statistically independent events before accounting for detector efficiency, source averaging, nuisance parameters, and the fact that flavor projection may not saturate the QFI. This is a necessary local quantum-limit count under the stated coherent-state and known-nuisance assumptions, not an achievable forecast for a solar detector. Complete phase averaging makes the temperature FI vanish for the purely longitudinal channel considered above.

The key distinction is now explicit. The index $\ident$ in Fig.~\ref{fig:identmap} answers whether two thermal parameters can be distinguished \emph{in principle}. The QSNR in Fig.~\ref{fig:solarqsnr} answers whether the corresponding temperature imprint is large enough to measure. A well-conditioned QFIM can still be useless if all of its eigenvalues are microscopically tiny.

\section{Scope of the transverse-channel and two-probe scale argument}

Near resonance, the transverse term in Eq.~\eqref{eq:hfluc} becomes dominant. This is important because it opens finite-frequency transition channels and may improve multiparameter conditioning. Recent nonequilibrium thermometry shows that noncommuting couplings can substantially modify QFI when dephasing and dissipation act on comparable dynamical timescales \cite{Aiache2025}. In the Sun, however, the total microscopic fluctuation power is controlled by the same weak potential. The MSW rotation redistributes this power because
\begin{equation}
\cos^22\theta_m+\sin^22\theta_m=1;
\end{equation}
however, it does not create a new interaction scale.

The rotation identity alone does not establish an orientation-independent bound on information; the spectrum and its temperature derivatives at zero and finite frequencies must also be controlled. Within an explicitly broadband Debye-cell model with logarithmic rate derivatives of order unity, both channels remain perturbatively weak over a solar-radius baseline. A general microscopic no-go theorem would require bounding the trajectory spectrum, the associated temperature derivatives, and any temperature-dependent coherent phase.

Indeed, at 10 MeV, the cell estimate gives $\Omega_m\tau_D/\hbar\simeq1.26\times10^{-15}$, so its exponential-correlation spectrum is effectively flat across the matter gap. Transverse noise is not spectrally suppressed and becomes dominant near local resonance. The longitudinal resonance dip therefore does not bound the full-channel information. Two-probe benchmarks can redistribute information under fixed-resource assumptions, but they do not by themselves prove an exclusion for an arbitrary number of many-probe strategies.

We therefore retain this section only as a conditional argument for the scale. No figure in this manuscript is interpreted as a calculation of the full longitudinal-plus-transverse solar sensitivity. A quantitative transverse result requires evaluating the finite-frequency trajectory spectrum in Eq.~\eqref{eq:traj} and propagating the resulting population dynamics through production and detection averaging.

\section{Discussion}

\subsection{Contribution and relation to established estimation results}

The rank-one outer-product algebra, Fisher-matrix addition for independent samples, and correlated-dephasing Bell-sector structure are established estimation-theory ingredients rather than new general theorems \cite{Tamascelli2020,AsjadTekluParis2023,Lu2026,LoulijatTeklu2026}. The contribution claimed here is their conditional application to MSW-dependent thermal-profile weights, together with a physically constrained separation between identifiability and the weak solar fluctuation scale. We make no broad priority claim for energy-resolved rank restoration or Bell-sector thermometry.

Within the restricted two-zone neutrino model, the first useful consequence is the rank statement in Eq.~\eqref{eq:rankone}. It sharpens the distinction between poor precision and structural non-identifiability. If two temperatures enter the state only through a single scalar dephasing exponent, no increase in statistics can distinguish them. This is a structural inverse-problem obstruction. The relevant remedy is not ``more QFI'' in the same channel, but rather an additional channel with a different sensitivity direction.

The neutrino-specific step is that the MSW energy dependence can supply distinct sensitivity directions without externally engineering the probe. Because $\theta_m(E,n_e)$ changes differently in different density regions, energy diversity acts as a form of natural control. In this sense, solar-neutrino spectroscopy is also a form of sensing-geometry engineering. The usefulness of an energy bin is determined not only by its event rate or oscillation amplitude, but by whether its thermal sensitivity vector adds a new direction to the Fisher span.

The two-probe extension then illustrates, rather than newly establishes, the separation between common-mode and differential dephasing sectors. The $\Phi$ and $\Psi$ sectors respond to sum and difference phases, respectively. If a correlated-neutrino source or a faithful quantum simulator were available, the pair could estimate not only an effective temperature parameter but also a spatial correlation length. This is closer to environmental noise spectroscopy than to a universal entanglement advantage.

\subsection{Relation to inverse solar-neutrino studies}

The inverse perspective is consistent with recent work that constrains $n_e(r)$ directly from flavor data rather than taking it only as a forward-model input \cite{LaberSmith2024}. The present problem is more difficult because the leading MSW Hamiltonian is directly sensitive to $n_e$, whereas the direct temperature dependence enters only through much weaker fluctuation and finite-temperature channels. Consequently, a realistic likelihood must treat density, composition, oscillation parameters, flux normalizations, and detector response as nuisance parameters. A single-temperature QFI, computed with all other quantities held fixed, is not a realistic solar uncertainty.

Borexino's pp-chain measurements and CNO-neutrino evidence provide complementary observational context \cite{Borexino2018,BorexinoCNO2020}. Inferences from production fluxes involve the solar nuclear-reaction model, the temperature profile, and the composition. They must be distinguished from the much smaller propagation-induced fluctuation signal studied here; these measurements do not constitute observations of thermal neutrino dephasing.

A useful future parametrization is
\begin{equation}
T(r)=T_{\rm SSM}(r)\left[1+a_1B_1(r)+a_2B_2(r)+\cdots\right].
\label{eq:profile}
\end{equation}
The Fisher matrix for the coefficients $a_k$ can then be diagonalized. Its eigenvectors identify the thermal profile combinations to which the neutrino data are actually sensitive. This provides a mathematically clean route from the present two-zone \revision{criterion} to continuous thermal tomography.

\subsection{Physical meaning of the conditional solar benchmark}

A negative solar result is not a failure of the formalism. Within the chosen cell model, the small signal follows directly from the very small weak-potential noise scale; a response-theory derivation is still required before it can be promoted to a quantitatively controlled Standard-Model bound. Large mesoscopic density structures may alter MSW evolution much more strongly than equilibrium thermal fluctuations, as known from stochastic-Sun studies \cite{LoretiBalantekin1994,Nunokawa1996,BurgessMichaud1997}. However, such structures are not thermometers unless their amplitudes and correlation lengths are independently related to thermodynamic temperature. The distinction between ``density-noise spectroscopy'' and ``temperature inference'' must remain explicit.

Similarly, finite-temperature refractive corrections exist in thermal field theory \cite{NotzoldRaffelt1988,Tututi2002}, but at solar-core temperatures they are far below the ordinary density-induced MSW potential. The mean Hamiltonian is therefore essentially a density probe. A small thermal refractive correction alone does not quantify its QFI: its temperature derivative and accumulated phase must also be calculated. Conventional temperature inference through neutrino production rates or solar-model relations is outside the propagation-only parameter model used here.

\section{Limitations and controlled extensions}
\label{sec:limitations}

\paragraph{Two-flavor approximation.}
The two-flavor solar sector isolates the $\Delta m^2_{21}$ and $\theta_{12}$ physics responsible for the dominant MSW effect. A precision analysis should embed the open-system terms in the full three-flavor Hamiltonian and verify that $\theta_{13}$ corrections do not alter the qualitative conclusions. Because the quoted longitudinal Debye-cell QSNR is extremely small, moderate three-flavor corrections to the coherent MSW sector would not by themselves make that specific benchmark operationally large; a full three-flavor open-system calculation is nevertheless required before extending this statement beyond the benchmark.

\paragraph{Two-zone model.}
The benchmark zones are not literal thermodynamic reservoirs. They discretize a continuous $T(r),n_e(r)$ profile so that the identifiability theorem is transparent. The next step is a radial Liouvillian
\begin{equation}
\frac{d\rho}{dr}=\mathcal L[r,E,T(r),n_e(r)]\rho
\end{equation}
combined with standard solar production distributions.

\paragraph{Microscopic bath spectrum.}
The Debye-cell construction is a heuristic order-of-magnitude estimate, not a substitute for $S_{nn}(q,\omega)$ from plasma response theory. A quantitative microscopic extension should evaluate $\chi^R_{nn}$ in RPA or kinetic theory, include ions and screening, and fold the result with the neutrino wave-packet trajectory. This is the most important quantitative improvement to the current benchmark, together with a finite-temperature evaluation of the collisional damping term of Sec.~\ref{sec:debye}.

\paragraph{Markov and secular approximations.}
The completely positive semigroup framework underlying the generator is established by Gorini, Kossakowski, and Sudarshan and by Lindblad \cite{Gorini1976,Lindblad1976}. It constrains the allowed reduced dynamics but does not determine a microscopic plasma rate. Treatments beyond the Markov approximation require additional dynamical information \cite{Breuer2016}.

Equation~\eqref{eq:master} assumes weak coupling and appropriate separation of bath and system timescales. These assumptions must be checked by comparing the plasma correlation time, $\Omega_m^{-1}$, and the radial variation scale of the solar medium. If memory effects are relevant, a non-Markovian treatment should replace the secular Lindblad equation.

\paragraph{Adiabaticity and source averaging.}
The instantaneous matter basis is always a valid representation; neglecting its connection term requires sufficiently adiabatic evolution. Solar neutrinos are also produced over an extended region. Averaging over the production distribution and the detector energy resolution can wash out phase information and must not be mistaken for thermal decoherence.

\paragraph{Wave-packet separation.}
The wave-packet treatment supplies the coherence conditions for neutrino oscillations \cite{GiuntiKim1998}. Recent information-theoretic analysis likewise distinguishes incoherent solar ensembles from coherent reactor propagation \cite{ChundawatLi2026}. That \revision{work} concerns oscillation-parameter information; the loss of temperature information under complete phase averaging in our restricted dephasing model follows from the temperature-independent averaged state, not from importing its parameter-specific QFI results.

Kinematical separation of mass wave packets provides an independent source of coherence loss. Any data analysis must model it separately from environment-induced dephasing, since only the latter is tied to the medium fluctuation spectrum.

\paragraph{Two-neutrino entanglement.}
Prepared Bell pairs of solar neutrinos are not an available astrophysical resource. The two-qubit correlated-bath calculations are therefore metrological benchmarks. The experimentally closer resource is energy-resolved sampling by many independent neutrinos.

\paragraph{Multiparameter compatibility.}
A nonsingular QFIM does not by itself guarantee simultaneous saturation. However, the scalar-dephasing model has $L_i=(\partial_i\Gamma)\,L_\Gamma$ within each energy class. These SLDs commute, and the SLDs of the product state over independent labeled samples are sums of such local operators acting on different factors, so they also commute. This establishes local SLD compatibility for that restricted model, but not the feasibility of the corresponding measurement in a solar detector. Adding populations, phases, or nuisance parameters requires a fresh analysis \cite{Ragy2016,Liu2020,AsjadTekluParis2023}.

\paragraph{Beyond the Sun.}
The equilibrium Sun is intentionally a low-complexity, low-sensitivity benchmark. Core-collapse supernovae, proto-neutron stars, and neutron-star merger remnants are hotter, denser, collisional, and nonequilibrium. In those media, finite-temperature self-energy, scattering, damping, neutrino--neutrino refraction, and time-dependent fluctuations are stronger. The present chain---microscopic response, open-system propagation, QFIM, and operational readout---provides a consistency template for studying whether neutrino thermometry becomes viable there.

\section{Conclusions}

We have formulated solar-neutrino quantum thermometry as a multiparameter open-system inference problem. The mean electron density controls coherent MSW propagation, while temperature enters through the microscopic fluctuation spectrum. On the matter basis, density noise automatically generates a noncommuting longitudinal--transverse coupling whose orientation and transition frequency are controlled by neutrino energy.

The central analytical result is an identifiability criterion for the longitudinal two-zone problem, obtained by applying standard Fisher-matrix rank arguments to MSW-dependent thermal weights. If two temperatures enter a single-energy-output state only through a single accumulated dephasing exponent, then the QFIM is an outer product and has rank 1. Separate estimation of $T_1$ and $T_2$ is then impossible regardless of statistics. Multiple energies can restore full rank when their MSW-dependent thermal-sensitivity vectors are noncollinear. The two-zone benchmark demonstrates a strong contrast between well-conditioned pairs in the MSW-sensitive region and nearly redundant high-energy pairs.

A genuine two-probe extension adds spatial information. Local and common baths produce different metrological structures, and Bell sectors separate common-mode from differential fluctuations, providing a theoretical route to joint temperature--correlation-length sensing. This extension should be understood as environmental correlation spectroscopy rather than as a universal advantage in entanglement.

Finally, a heuristic electron-only Debye-cell model yields a conditional longitudinal coherence decay scale on the order of $10^{-12}\,\mathrm{s}^{-1}$. For a coherent flavor state at 10 MeV in a uniform central medium over one solar radius, the QSNR is approximately $3.1\times10^{-13}$, giving an ideal local event lower bound of roughly $3.3\times10^{14}$ for 10\% relative precision. These values do not constitute a general microscopic no-go theorem or a detector forecast. In particular, complete phase averaging removes a temperature imprint carried solely by longitudinal coherence damping, and incoherent $\nu$--$e$ collisions provide a larger, although only weakly temperature-dependent, decoherence rate that remains to be bounded.

The established results are conditional identifiability criteria and reproducible illustrative scales. A quantitative solar inference calculation must determine the trajectory-sampled fluctuation spectrum, include transverse dynamics and production/detection averaging, and evaluate the resulting event information with nuisance parameters. Energy diversity can remove degeneracies only in temperature-sensitive channels that survive those operations.
\appendix
\section{Derivation of the rank-one QFIM}
\label{app:rankone}
This appendix derives Eqs.~\eqref{eq:outer} and \eqref{eq:rankone} step by step and states the conditions under which they hold.

\subsection{General argument}

\paragraph{Assumptions.} At fixed neutrino energy $E$, the propagated state is assumed to depend on the thermal parameters $\vartheta_j=\ln(T_j/T_{j,0})$ only through the accumulated exponent,
\begin{equation}
\rho=\rho\big(\Gamma(\vartheta_1,\vartheta_2)\big),\qquad
\Gamma=\Gamma_1(E,T_1)+\Gamma_2(E,T_2).
\label{eq:app_assump}
\end{equation}
This requires temperature-independent preparation, mean Hamiltonian (and hence $\theta_m$ and the phase $\phi$), and Lamb shift. It also requires negligible transverse rates and statistically independent zone noise, so that the exponents add.

\paragraph{Step 1: chain rule.} Because $\rho$ depends on $\vartheta_i$ only through $\Gamma$,
\begin{equation}
\partial_{\vartheta_i}\rho=\rho'(\Gamma)\,g_i,\qquad
\rho'\equiv\frac{d\rho}{d\Gamma},\qquad
g_i\equiv\partial_{\vartheta_i}\Gamma .
\label{eq:app_chain}
\end{equation}

\paragraph{Step 2: proportional SLDs.} Define $L_\Gamma$ through $\rho'=\tfrac12(L_\Gamma\rho+\rho L_\Gamma)$. Since the SLD equation is linear in $L$, the operator
\begin{equation}
L_i=g_i\,L_\Gamma
\label{eq:app_sld}
\end{equation}
satisfies $\partial_i\rho=\tfrac12(L_i\rho+\rho L_i)$. For $\Gamma>0$ the Bloch vector obeys $|\bm r|<1$, so $\rho$ is full rank and this solution is unique. All SLDs are multiples of a single operator and therefore commute, which is the origin of the compatibility statement in Sec.~\ref{sec:limitations}.

\paragraph{Step 3: outer-product structure.} Inserting Eq.~\eqref{eq:app_sld} into the definition of the QFIM gives
\begin{equation}
\QFI_{ij}=\tfrac12\Tr\big[\rho\{L_i,L_j\}\big]
=g_ig_j\,\Tr\big[\rho L_\Gamma^2\big]\equiv K\,g_ig_j ,
\label{eq:app_outer}
\end{equation}
which is Eq.~\eqref{eq:outer}. The scalar $K=\QFI_{\Gamma\Gamma}\geq0$ is the single-parameter QFI for $\Gamma$ itself.

\paragraph{Step 4: vanishing determinant.} For two parameters,
\begin{equation}
\QFI^{(E)}=K\begin{pmatrix}g_1^2&g_1g_2\\ g_1g_2&g_2^2\end{pmatrix},
\qquad
\det\QFI^{(E)}=K^2\left(g_1^2g_2^2-g_1^2g_2^2\right)=0 ,
\end{equation}
which is Eq.~\eqref{eq:rankone}. Equivalently, the vector $\bm w=(g_2,-g_1)^T$ satisfies
\begin{equation}
\QFI^{(E)}\bm w=K\,\bm g\,(\bm g\cdot\bm w)=0 ,
\end{equation}
so $\QFI^{(E)}$ has a null direction. Its rank is exactly one when $K>0$ and $\bm g\neq0$. The same argument applies to any number of parameters, since $\bm g\bm g^T$ has rank at most one.

\paragraph{Step 5: independence of statistics.} For $M$ independent events the total QFIM is $M\QFI^{(E)}$, and $\det(M\QFI^{(E)})=M^2\det\QFI^{(E)}=0$. A displacement $\delta\bm\vartheta\propto\bm w$ gives $\delta\Gamma=\bm g\cdot\delta\bm\vartheta=0$ and therefore leaves $\rho$ unchanged to first order. No measurement on any number of copies can resolve it; only the combination $\bm g\cdot\delta\bm\vartheta$ is constrained.

\subsection{Explicit prefactor for qubit states}
Consider an equatorial qubit undergoing pure dephasing,
\begin{equation}
\rho=\frac{1}{2}\begin{pmatrix}1&e^{-\Gamma-i\phi}\\e^{-\Gamma+i\phi}&1\end{pmatrix},
\end{equation}
where the phase is independent of the thermal parameters. The Bloch vector is $\bm r=e^{-\Gamma}(\cos\phi,\sin\phi,0)$.  For parameters $\lambda_i$,
\begin{equation}
\QFI_{ij}=\partial_i\bm r\cdot\partial_j\bm r+\frac{(\bm r\cdot\partial_i\bm r)(\bm r\cdot\partial_j\bm r)}{1-|\bm r|^2}.
\end{equation}
Since $\partial_i\bm r=-(\partial_i\Gamma)\bm r$,
\begin{equation}
\QFI_{ij}=\frac{\partial_i\Gamma\,\partial_j\Gamma}{e^{2\Gamma}-1}.
\end{equation}
The matrix is the outer product of $\nabla\Gamma$ with itself. It therefore has rank at most one for any number of parameters whose only imprint is through $\Gamma$, and exactly one when the scalar response and gradient are nonzero.

For an electron-flavor preparation in a constant-density medium, the Bloch vector instead has fixed longitudinal component $\cos2\theta_m$ and transverse length $e^{-\Gamma}\sin2\theta_m$. 
Writing $s=\sin2\theta_m$ and $c=\cos2\theta_m$,
\begin{equation}
\bm r=\big(s\,e^{-\Gamma}\cos\phi,\; s\,e^{-\Gamma}\sin\phi,\; c\big).
\end{equation}
Only the transverse components depend on $\Gamma$, so
\begin{equation}
\partial_i\bm r\cdot\partial_j\bm r=s^2e^{-2\Gamma}g_ig_j,\qquad
\bm r\cdot\partial_i\bm r=-s^2e^{-2\Gamma}g_i,\qquad
1-|\bm r|^2=s^2\left(1-e^{-2\Gamma}\right),
\end{equation}
with $g_i=\partial_i\Gamma$. Substituting into the Bloch formula,
\begin{equation}
\QFI_{ij}=s^2e^{-2\Gamma}g_ig_j\left[1+\frac{e^{-2\Gamma}}{1-e^{-2\Gamma}}\right],
\end{equation}
and therefore
\begin{equation}
\QFI_{ij}=\frac{\sin^22\theta_m}{e^{2\Gamma}-1}\partial_i\Gamma\,\partial_j\Gamma.
\end{equation}
This establishes the prefactor used in Eq.~\eqref{eq:qt} when the mixing angle and phase are independent of the parameters of interest. These expressions apply in the regular interior; boundary points must be evaluated by an appropriate limit.
Comparison with Eq.~\eqref{eq:app_outer} identifies $K=\sin^22\theta_m/(e^{2\Gamma}-1)$, with $\sin^22\theta_m\to1$ for an equatorial preparation. In the solar regime $\Gamma\ll1$, this gives $K\simeq\sin^22\theta_m/(2\Gamma)$.

\subsection{Sensitivity vector under Debye scaling}

At fixed density, the Debye-cell estimate of Sec.~\ref{sec:debye} gives $\gamma_{\phi,j}\propto\cos^22\theta_{mj}\,n_{ej}^2/T_j$, so $\Gamma_j\propto T_j^{-1}$. By zone additivity, $\Gamma_j$ depends only on $T_j$, and
\begin{equation}
g_j=T_j\,\partial_{T_j}\Gamma_j=-\Gamma_j,
\qquad
\bm g(E)=-\big(\Gamma_1(E),\,\Gamma_2(E)\big).
\end{equation}
The direction of $\bm g$ is therefore fixed by the ratio $\Gamma_2/\Gamma_1$, which is the angle $\alpha(E)$ shown in Fig.~\ref{fig:direction}. The weights $\cos^22\theta_{mj}(E)$ make this direction energy dependent, and Sec.~\ref{sec:energy} uses this to restore full rank with two energies.

\subsection{Conditions for validity}

The rank-one result relies on Eq.~\eqref{eq:app_assump}. It fails as soon as temperature enters $\rho$ in any other way: through a thermal refractive contribution to $\phi$, through temperature-dependent populations generated by transverse transitions, or through a temperature-dependent $\theta_m$. In each case $\partial_i\rho$ acquires a component that is not parallel to $\rho'(\Gamma)$, and the outer-product structure of Eq.~\eqref{eq:app_outer} is lost. At the boundary $\Gamma=0$, where the state is pure and $K$ diverges, the expressions must be taken as a limit.

\section{Two-energy determinant}

For two independent energy classes,
\begin{equation}
\QFI=K_1\bm g_1\bm g_1^T+K_2\bm g_2\bm g_2^T.
\end{equation}
In two dimensions, direct expansion gives
\begin{equation}
\det\QFI=K_1K_2(g_{11}g_{22}-g_{12}g_{21})^2.
\end{equation}
Event multiplicities multiply the corresponding $K_a$ factors. The determinant is positive if and only if the two sensitivity vectors are noncollinear.
\section*{Acknowledgments}
The research conducted by BT was supported by Khalifa University of Science and Technology under Project No. KU-INT-RIG-2024-8474000739.

\section*{Data Availability} The numerical data and plotting code that support the findings of this article are available from the corresponding author upon reasonable request.


\end{document}